\documentclass[12pt]{article}
\usepackage{epsfig}
\usepackage{amsfonts}
\newcommand{\be}{\begin{equation}}
\newcommand{\ee}{\end{equation}}
\newcommand{\bea}{\begin{eqnarray}}
\newcommand{\eea}{\end{eqnarray}}
\def\half{\frac{1}{2}}
\begin{document}

\begin{titlepage}
\begin{flushright}
hep-th/0411187
\end{flushright}
\vskip 1.5in
\begin{center}
{\bf\Large{Microscopic Entropy of the Black Ring}}
\vskip 0.5in
{Michelle Cyrier$^{a}$, Monica Guica$^{a}$,
David Mateos$^{b}$ and Andrew Strominger$^{a}$}
\vskip 0.3in
{\small {\textit{$^{a}$Jefferson Physical Laboratory, Harvard University, Cambridge MA 02138, USA}
}}
\\ {\small{
\textit{ $^b$Perimeter Institute for Theoretical Physics, Waterloo, Ontario N2J 2Y5, Canada}}}

\end{center}
\vskip 0.5in

\baselineskip 16pt
\date{}

\begin{abstract}
A surprising new seven-parameter supersymmetric black ring solution of
five-dimensional supergravity has recently been discovered. In this
paper, M-theory is used to give an exact microscopic accounting of
its entropy.

\end{abstract}
\end{titlepage}
\vfill\eject

A supersymmetric black ring has been recently discovered
\cite{littleeemr}. This is an asymptotically flat black hole
solution of five-dimensional supergravity whose event horizon has
toroidal topology $S^1 \times S^2$. A more general seven-parameter
supersymmetric black ring solution was found in \cite{bigeemr,GG2}.
Previous studies of non-BPS or singular black rings include
\cite{previous}. Concentric black ring solutions have been found in 
\cite{GG2,GG1}. 

In M-theory the solutions of \cite{littleeemr,bigeemr} correspond to
supertubes \cite{mt}, namely to configurations with M5-branes
wrapped around four-cycles of a six-torus and the fifth brane
dimension forming a circle, or ring,  stabilized by angular momentum
in five-dimensional space-time. There is also M2-brane charge
density distributed around the ring. The brane wrappings around the
different cycles are summarized by the array

\begin{center}
\begin{tabular}{ccccccccccc}
$q_1$ & M2:& & 1&2&-&-&-&-&-\\
$q_2$ & M2:& & -&-&3&4&-&-&-\\
$q_3$ & M2:& & -&-&-&-&5&6&-\\
$p^1$ & M5:& & -&-&3&4&5&6&$\psi$ \\
$p^2$ & M5:& & 1&2&-&-&5&6&$\psi$ \\
$p^3$ & M5:& & 1&2&3&4&-&-&$\psi$
\end{tabular}
\end{center}
where $q_A$ and $p^A$, $A=\{1,2,3\}$, are the numbers of M2-branes
and M5-branes, respectively, wrapping each cycle, and $\psi$ is the
five-dimensional angular coordinate around the ring. The solution of
the five-dimensional effective supergravity theory is reproduced in
the appendix of this paper. It is characterized by seven independent
parameters, which we choose to be the set of six brane numbers
listed above, together with the angular momentum $J_{\psi}$ around
the ring.

The Bekenstein-Hawking entropy, $S= \mathcal{A}_{horizon} / (4G)$, of this
supergravity solution is given in \cite{bigeemr} in the form
\be
S_{BR} = 2 \pi \sqrt{q_1 q_2q_3 - k_1 k_2 k_3 - J_{\phi}^2 -
D(J_{\psi} - J_{\phi}) } \,,
\ee
where $k_1 \equiv q_1 -p^2 p^3$ (and similarly for permutations of
1, 2 and 3), $D\equiv p^1p^2p^3$ and $J_\phi$ is the angular
momentum in the plane orthogonal to the ring. Replacing $J_{\phi}$
by its expression (\ref{jphi}) in terms of $q_A$, $p^A$, we get
\be
S_{BR} = 2 \pi \sqrt{ \frac{D^2}{4} - D J_{\psi} -
\frac{1}{4} \Big( (p^1q_1)^2  + (p^2q_2)^2  + (p^3q_3)^2  \Big) +
\half D \left( \frac{q_1 q_2}{p^3} + \frac{q_2 q_3}{p^1} +
\frac{q_1 q_3}{p^2} \right) } \label{brentropy}
\ee
where we have written everything in terms of the seven independent
integers $q_A,p^A,J_{\psi}$. We wish to compare this expression with
the formula we get from the microscopic computation.~\footnote{A
previous discussion of this problem can be found in
\cite{bena}.}

Our starting point is reference  \cite{msw}, where a derivation of
the entropy of a four-dimensional black hole in a Calabi-Yau$\times S^1$
compactification of M-theory was given.\footnote{Although proper
$SU(3)$ holonomy was assumed in \cite{msw}, the  
results quoted here are valid for a six-torus. } The
construction involves an M5-brane wrapped about
$\sum_Ap^A \Sigma_A$, where $\Sigma_A$ is an integral basis of four
cycles in the Calabi-Yau. The resulting string in five dimensions is
a chiral (0,4) CFT, whose left-moving central charge is given to
leading order by $c_{L}=6D$, where
\be
D = D_{ABC} p^{A}p^{B}p^{C}
\ee
and $D_{ABC}$ are the triple intersection numbers of the $\Sigma_A$.
For the case of a six-torus, relevant to the solutions of
\cite{littleeemr,bigeemr}, $D_{ABC}$ is equal to $1/6$ if $(ABC)$ form
a permutation of $(123)$ and zero otherwise. Therefore  we can write
\be
c_L = 6 D= 6 p^1 p^2 p^3 .
\ee
Membrane charge arises as the momentum zero modes $q_A$ of a Narain
lattice of scalars in the CFT.  The left-moving oscillator number,
denoted  $\hat q_0$ in \cite{msw}, was shown to be related to the
left-moving momentum $q_0$ by\footnote{The last zero-point term was
omitted in \cite{msw} because it was subleading in the parameter
  range considered therein.}
\begin{equation}
\hat{q}_{0}=q_{0} +\frac{1}{12}D^{AB}q_{A}q_{B} + \frac{c_L}{24} \,,
\end{equation}
where $D^{AB}$ is the inverse of the matrix $D_{AB}$, defined as
\be
D_{AB} = D_{ABC}p^C = \frac{1}{6} \left( \begin{array}{ccc}0&p^3&p^2 \\
p^3 & 0 & p^1 \\ p^2 & p^1 &0 \end{array} \right).
\ee
The second term in $\hat q_0$ comes from the momentum carried by the
Narain scalar zero modes, while the last term is the usual
oscillator zero-point
contribution to the momentum.
The entropy of excited states with left oscillator number $\hat q_0$ is
then given by the Cardy formula
\be
S = 2\pi \sqrt{ c_{L} \hat{q}_{0} /6} \label{micro} \,,
\ee
where $\hat{q}_0$ is the left moving momentum available to be distributed among the
oscillators.

In \cite{msw}, a further $S^1$ compactification from five to four
dimensions was performed, and four-dimensional black holes were made
by wrapping the string around this $S^1$. The resulting
macroscopic entropy agreed with the microscopic result following
from equation (7).  In the present context we do not wish to
compactify to four dimensions. However we can still put the same
(0,4) CFT on a circle in five dimensions by simply tying up the ends
of the string, rather than winding it around the $S^1$. The string is
now stabilized dynamically by the angular momentum $J_\psi$. Hence
we are proposing that the microscopic description of the black ring
is simply as a ring of the same CFT encountered in \cite{msw}.

This can be checked by computing the microscopic entropy. The
left-moving momentum $q_0$ should equal $J_{\psi}$ up to a sign,
since this is the momentum around the ring, and both are integrally
quantized. Taking the sign such that $q_0=-J_\psi$ we find
\be
\hat{q}_0 = - J_{\psi} + \frac{D}{4} +
\half \left( \frac{q_1 q_2}{p^3} + \frac{q_2 q_3}{p^1} +
\frac{q_1 q_3}{p^2} \right) - \frac{1}{4 D} \Big(
(p^1  q_1)^2 + (p^2 q_2)^2 + (p^3 q_3)^2 \Big) \,.
\ee
Substituting this expression into the Cardy formula (\ref{micro})
for the entropy and comparing with the supergravity result
(\ref{brentropy}) we find perfect agreement.

In conclusion, the microscopic description of the black ring
solution of \cite{littleeemr,bigeemr} is that of a ring of the same
Calabi-Yau-wrapped M5-brane (0,4) CFT encountered  in \cite{msw}.

\section*{Appendix}
The five-dimensional metric has the form
\be ds^2 = -f^2(dt + \omega)^2 + f^{-1}
\frac{R^2}{(x-y)^2} \left[
\frac{dy^2}{y^2-1} + (y^2-1)d\psi^2 + \frac{dx^2}{1-x^2} +
(1-x^2)d\phi^2 \right]
\ee
where $f^{-1}=(H_1 H_2 H_3)^{1/3}$,
\be
H_1 = 1 +
\Big(\frac{4G}{\pi}\Big)^{2/3}
\frac{q_1 - p_2 p_3}{2R^2}(x-y) -
\Big(\frac{4G}{\pi}\Big)^{2/3}
\frac{p_2 p_3}{4R^2}(x^2 - y^2)
\ee
and $H_2$ and $H_3$ are given by obvious permutations.

The coordinates have ranges $x \in [-1,1], y \in [-\infty ,-1],
\phi,\psi \in [0,2\pi] $ and asymptotic infinity lies at
$x\rightarrow y \rightarrow -1$. Further, for the black ring
solution we have $\omega = \omega_{\phi} d\phi + \omega_{\psi}
d\psi$, where
\bea
\omega_{\phi} & = & - \frac{G}{2\pi R^2 } (1-x^2) [ p^1 q_1 +
p^2 q_2 + p^3 q_3 - D (3+x+y)] \,, \\ \nonumber
 \omega_{\psi} & = &
\half \left(\frac{4G}{\pi} \right)^{\frac{1}{3}}(1+y) - \frac{G}{2\pi
R^2 } (y^2-1) [ p^1 q_1 + p^2 q_2 + p^3 q_3 - D (3+x+y)] \,,
 \eea
with $D=p^1p^2p^3$. The solution possesses two angular momenta as
measured at infinity. In terms of the brane numbers and the radius
$R$ they are given by
\bea
J_{\phi} & = & \half(p^1 q_1 + p^2 q_2 + p^3 q_3 - D)  \,, \\
\label{jphi} J_{\psi} & = & \left(\frac{\pi}{4G}
\right)^{\frac{2}{3}} R^2 (p^1 + p^2 + p^3)
+ \half (p^1 q_1 + p^2 q_2 + p^3 q_3 - D) \,.
\eea

Note that in terms of the charges used in the first reference in
\cite{bigeemr} we have
\be
q_i = \left(\frac{\pi}{4G}\right)^{\frac{2}{3}} Q_i^{EEMR}
\,\,,\qquad p^i = \left(\frac{\pi}{4G}\right)^{\frac{1}{3}}
q_i^{EEMR} \,.
\ee
 \medskip
\section*{Acknowledgments}
\noindent
We thank H. Elvang, R. Emparan, D. Jafferis, J. Maldacena, D. Marolf, H. Reall
and H. Verlinde for valuable discussions. This work was supported in part by DOE
grant DE-FG02-91ER40654.


 \end{document}